\documentstyle[12pt,a4]{article}
\begin{document}
\begin{flushright}
PRA-HEP-92/18\\
\end{flushright}
\vspace{5ex}
\begin{center}
{\LARGE \bf
Quantum braided groups}\\[0.8in]
{\sc  Ladislav   Hlavat\'{y}}\\[0.1in]
{\it Institute of Physics, Czechoslovak Academy of Sciences}\\
 {\it and}
\\ {\it Department  of  Physics,}
\\ {\it  Faculty  of  Nuclear  Sciences  and
Physical Engineering}
\\ {\it Prague, Czechoslovakia}\footnote{Postal address: B\v{r}ehov\'{a} 7, 110
Prague 1, Czechoslovakia.
 E-mail: HLAVATY@FJFI.CVUT.CS}\\
\vspace{.8in}
{\bf Abstract}\\
\end{center}
\begin{quotation}
A new type of algebras that represent a generalization of both quantum
groups and braided groups is defined. These algebras are given by  a  pair
of solutions of the Yang--Baxter equation that  satisfy  some
additional conditions.  Several
examples are presented.
\end{quotation}
\newpage

\section{Introduction -- quantum groups and braided groups}

Matrix groups like $GL(n), SO(n)$ e.t.c. were generalized in  two
ways  recently.  Both  are based on deformation of  the  algebra   of
functions  on  the  groups generated   by
coordinate functions $T_{i} ^{j}$ that  commute
\begin{equation}
T_{i} ^{j}T_{k} ^{l}  = T_{k} ^{l}T_{i} ^{j}
\ \ \Leftrightarrow \ \ T_{1}T_{2} = T_{2}T_{1}
\label{tt}
\end{equation}

In the quantum groups \cite{Drinfeldqg,FRT}
these commutation relations are modified by a matrix $R = \{
R_{ij} ^{kl}  \}$  so
that the functions do not commute but satisfy the relations
\begin{equation}
R_{12}T_{1}T_{2} = T_{2}T_{1}R_{12}
\label{rtt}
\end{equation}
In this relation the elements of matrix $R$ are numbers but  the
 matrix $T=\{T_i^j\}$ is formed  by  generally  noncommuting
elements of an algebra.

Another type of deformation of the relations  (\ref{tt})  represent  the
so called  braided groups
\cite{Majidjmp}
defined by the relation
\begin{equation}
T_{1}Z_{12}T_{2}Z_{12} ^{-1} = Z_{21} ^{-1} T_{2}Z_{21}T_{1}
\label{tztz}
\end{equation}
where $Z$ is again a matrix $\{ Z_{ij} ^{kl} \} $ with number elements.

The quantum groups appeared to be  hidden  symmetries  of  many  physical
models.
The  relevance    of    the    braided   groups    for    the
low--dimensional  quantum field theory  was
explained in \cite{Majidlmp}.

The goal of this paper is to define a concept that  unifies  both
the quantum groups and  braided  groups.  We  call  these  more  general
objects  {\em  quantum braided groups}.
Before doing that let us summarize the properties of  quantum
groups  and braided groups.

Both the algebras defined by
(\ref{rtt}) or
(\ref{tztz}) can  be  extended  to  bialgebras
with matrix coproduct and counit

\begin{equation}\Delta (T_{i}^{j}) := T_{i}^{k}\otimes T_{k}^{j},\ \epsilon(T
_{i}^{j}) := \delta_{i}^{j}.
\label{cop} \end{equation}
However, the tensor products of the algebras defined  by
the relations (\ref{rtt}) differ  from   those   defined   by
the  relations
(\ref{tztz}).
 The multiplication in  the  tensor  product
$A\otimes  A$  of  the   algebras   $A$   defined   by    the
relations
(\ref{rtt}) (corresponding to quantum groups) is
\[
m_{(A\otimes  A)}:  A\otimes  A\otimes  A\otimes  A   \rightarrow
A\otimes A  \]
\begin{equation}  m_{(A\otimes  A)}   =   (m\otimes   m)\circ
(id\otimes
 \tau_{23}\otimes id)
\end{equation}
where $m$ is the product in $A$ and $\tau_{23}$ is the transposition of
the second and third factor  in  $A\otimes  A\otimes  A\otimes  A$.
It is then easy to prove that $A$ is a bialgebra.

On the other hand the multiplication in the  tensor  product
$B\otimes  B$  of  the   algebras   $B$   defined   by    the
relations
(\ref{tztz}) (corresponding to braided groups) is more
complicated because instead
of the simple transposition $\tau$ a more general map $\psi:
B\otimes B  \rightarrow  B\otimes  B$  called  {\em braiding}
appears in the product \cite{Majidjmp}.
\[ m_{(B\otimes  B)}:  B\otimes  B\otimes  B\otimes  B   \rightarrow
B\otimes B  \]
\begin{equation}  m_{(B\otimes  B)} :=  (m\otimes  m)\circ
(id\otimes
 \psi_{23}\otimes id)
\label{prodintens}
\end{equation}
where $m$ is the product in $B$ and
\begin{equation}
\psi    (T_{i}     ^{m}     \otimes     T_{k}     ^{n})    :=
\psi_{ik}^{mn},_{rs}^{jl}(T_j^r \otimes T_l^s)
\label{psirel}
\end{equation}
where
\begin{equation}
\psi_{ik}^{mn},_{rs}^{jl}:=
Z_{id}^{aj}(Z^{-1})_{ar}^{lb}Z_{sb}^{cn} \tilde Z_{ck}^{md}
\label{psi}
\end{equation}
and $\tilde Z := ((Z^{t_2})^{-1})^{t_2}$.

To  prove  that  $B$  is
bialgebra namely that $\Delta$ and $\epsilon$  are  morphisms  of
the algebra $B$ and $B\otimes B$ is a bit more complicated  than  for
the quantum groups but there are no principal  problems.  The
identities
\begin{equation}
\tilde Z_{il}^{kn}Z_{kj}^{ml} =  Z_{il}^{kn}\tilde Z_{kj}^{ml} =
\delta_i^m \delta_j^n
\label{zztr2}
\end{equation}
which follow immediately from the  definition   of   $\tilde
Z$,  is
used for that.
If  antipodes on the bialgebras are defined we get Hopf
algebras.

\section{Quantum braided groups}

As mentioned in the beginning, our goal is  to  define  an  object
that will unify  the  properties  of  both  quantum  and  braided
groups or more precisely, that  will  contain  both  of  them  as
special cases. Prototypes for that are quantum supergroups.

The  supergroups
are    special  cases     of     the     braided groups      where     $
Z=\eta    :=$ \\
$diag(+, +, \ldots ,-,-,\ldots)$ and $\psi(x\otimes  y)  =
(-)^{|x||y|}y\otimes x$.
The
defining relations of a quantum supergroup can be written in a form that
reminds  (\ref{rtt})
but the supercommuting nature of its elements is expressed by
inclusion  of  the
 matrix $\eta  $ into the defining relation \cite{liao}
\begin{equation}
{\cal R}_{12}T_{1}\eta_{12}T_{2}\eta_{12}  =   \eta_{12}
T_{2}\eta_{12}T_{1}{\cal R}_{12}
\label{qsg}
\end{equation}

Comparing (\ref{rtt}), (\ref{tztz}) and (\ref{qsg}) leads us
quite naturally to the investigation of algebras  given
by a pair of $n^{2}\times n^{2}$ matrices $({\cal R},Z)$ that  define
 relations
\begin{equation}
{\cal R}_{12}T_{1}Z_{12}T_{2}Z_{12}   ^{-1}   =   Z_{21}   ^{-1}
T_{2}Z_{21}T_{1}{\cal R}_{12}             \label{qbgreln1}
\end{equation}
which include the cases of both quantum and braided groups.

The experience with the  quantum groups,  braided  groups,  and  quantum
supergroups  teaches  us  that  the
 matrices ${\cal  R}$  and  $Z$  cannot  be  arbitrary  but  will  be
restricted  by  conditions  of   Yang--Baxter   type.   These
conditions  follow  from
two possible ways to transpose  a  triple  of  generators  $T_{i}
^{j}$.
In  order that the relations (\ref{qbgreln1}) can  be  applied  we
shall consider triples of the form
\begin{equation}
T_{1}Z_{12}T_{2}Z_{12}^{-1}Z_{23}Z_{13}T_{3}Z_{13}^{-1}Z_{23}^{-1}
 \label{triple} \end{equation}
It can be transposed to an expression  with  the
transposed order of $T_{1},\ T_{2},\ T_{3}$
if matrices ${\cal R}$ and $Z$ are invertible and satisfy
\begin{equation} Z_{12}Z_{13}Z_{23}     =      Z_{23}Z_{13}Z_{12}
,\label{ybez}
\end{equation}
\begin{equation} {\cal R}_{12}Z_{23}Z_{13}= Z_{23}Z_{13}{\cal R}_{12}
,\label{trzz}
\end{equation}
\begin{equation} Z_{12}Z_{13}{\cal R}_{23}^{-1}Z_{32}^
{-1}  =
{\cal R}_{23}^{-1}Z_{32}^{-1}Z_{13}Z_{12}
\label{zztr}
\end{equation}
Under these conditions the  expression
 (\ref{triple}) can be transposed by two ways and we  require
that the results be equal
\[
{\cal R}^{-1}_{12}Z_{21}^{-1}{\cal
R}^{-1}_{13}Z_{31}^{-1}{\cal
R}^{-1}_{23}Z_{32}^{-1}T_{3}Z_{32}T_{2}Z_{31}Z_{21}T_1
{\cal    R}_{23}Z_{23}{\cal    R}_{13}Z_{13}{\cal     R}_{12}
Z_{13}^{-1}Z_{23}^{-1} =
\]
\begin{equation}
{\cal R}^{-1}_{23}Z_{32}^{-1}{\cal
R}^{-1}_{13}Z_{31}^{-1}{\cal
R}^{-1}_{12}Z_{21}^{-1}T_{3}Z_{32}T_{2}Z_{31}Z_{21}T_1
{\cal    R}_{12}Z_{12}{\cal    R}_{13}Z_{13}{\cal     R}_{23}
Z_{13}^{-1}Z_{12}^{-1}
\label{tripleeq}
\end{equation}
In order that the equation (\ref{tripleeq}) does not impose
additional  relations
for $T$ we require  that  the  matrix  ${\cal R}$
 satisfy the "braided Yang--Baxter equations"

\begin{equation}               {\cal       R}_{12}Z_{12}{\cal
R}_{13}Z_{13}{\cal R}_{23}Z_{23}  =
{\cal R}_{23}Z_{23}{\cal R}_{13}Z_{13}{\cal R}_{12}Z_{12}
\label{bybe}
\end{equation}
\begin{equation}
{\cal R}_{12}^{-1}Z_{21}^{-1}{\cal
R}_{13}^{-1}Z_{31}^{-1}{\cal R}_{23}^{-1}Z_{32}^{-1}  =
{\cal                            R}_{23}^{-1}Z_{32}^{-1}{\cal
R}_{13}^{-1}Z_{31}^{-1}{\cal R}_{12}^{-1}Z_{21}^{-1}
\label{bybe2}
\end{equation}

Introducing  $R :=  {\cal R}  Z$  we  immediately  see  that
(\ref{bybe}) is the ordinary Yang--Baxter equation (YBE) for $R$

\begin{equation} R_{12}R_{13}R_{23}     =      R_{23}R_{13}R_{12}
\label{yber}
\end{equation}
and the  equations  (\ref{trzz}) and (\ref{zztr}) can be
rewritten to simpler forms
\begin{equation} R_{12}Z_{13}Z_{23}     =      Z_{23}Z_{13}R_{12}
\label{rzz}
\end{equation}
\begin{equation} Z_{12}Z_{13}R_{23}     =      R_{23}Z_{13}Z_{12}
\label{zzr}
\end{equation}
The  equation  (\ref{bybe2})  is then satisfied   due   to \\
{\bf Lemma}: If $R$ and $Z$ are solutions of the   YBE   that
satisfy
(\ref{rzz})   and   (\ref{zzr})   then    $PZPRZ^{-1}$    and
$Z^{-1}RPZP$, where P is the permutation matrix  $P_{ij}^{kl}  =
\delta_i^l \delta_j^k$, are also solutions of the YBE.

Proof can be done by direct  check.  Let  us  note  that  the
condition $PZPZ^{-1} = 1$ required in  \cite{Reshetikhin}  is
not necessary here.

Conclusion then is that when we have a pair ($R,Z$) of solutions of the YBE
that satisfy the relations (\ref{rzz}),  (\ref{zzr}),  we
can define the algebra
\begin{equation}
B(R,Z) :=
C<T_{i}^{j}>_{i,j=1}^{n}/\{
R_{ab}^{cd}Z^{-1\ gh}_{cd}T_{g}^{i}Z_{ih}^{ml}T_{l}^{k}
-Z^{-1\
ji}_{ba}T_{j}^{g}Z_{gi}^{hc}T_{c}^{d}R_{dh}^{mk}\}_{a,b,m,k=1}^{n}
\label{dbgalgebra} \end{equation}
that do not impose additional relations
of cubic or higher degree. The compact form  of  the
relations
in (\ref{dbgalgebra}) is
\begin{equation}
R_{12}Z_{12} ^{-1}T_{1}Z_{12}T_{2}     =   Z_{21}   ^{-1}
T_{2}Z_{21}T_{1} R_{12}             \label{qbgreln}
\end{equation}
that is equivalent to (\ref{qbgreln1}).

One can show that
matrix coproduct
and counit (\ref{cop}) are morphisms of  $B(R,Z)$  into  $B(R,Z)\otimes
B(R,Z)$ where the product in $B(R,Z)\otimes B(R,Z)$ is defined by
(\ref{prodintens}), (\ref{psi}). To do that  one  must  prove
that
\begin{equation}
R_{12}Z_{12} ^{-1}\Delta (T_{1})Z_{12}\Delta (T_{2})     =   Z_{21}   ^{-1}
\Delta (T_{2})Z_{21}\Delta (T_{1}) R_{12}             \label{deltarln}
\end{equation}
which is simple  but  tedious  exercise  with  indices  where  the
identity (\ref{zztr2}) is used.
It means that the algebra (\ref{dbgalgebra}) can be  extended
to the bialgebra with the coproduct (\ref{cop}).
That enable
to   define   the   dual    algebra    of    regular
functionals $L^{\pm} = \{L^{\pm\ j}_i\}$ on
$B(R,Z)$.

 Let
\begin{equation}
<L^{\pm}_{1},T_{2}T_{3}\ldots T_{n}>        :=
{\cal R}^{\pm}_{12}{\cal R}^{\pm}_{13}
\ldots{\cal R}^{\pm}_{1n}
\end{equation}
where
\begin{equation}
{\cal R}^{+}_{12} := Z_{12}{\cal R}_{12} Z_{21} = Z_{12}R_{21},\ \
{\cal R}^{-}_{12} := {\cal R}^{-1} = Z_{12}R_{12}^{-1}
\end{equation}
and
\begin{equation}
 <ab,c> :=  <a\otimes b,\Delta (c)>             .
\end{equation}
The dual algebra is then characterized by the relations
\begin{equation}
{\cal R}_{21}L_{1}^{\epsilon}Z_{21}L_{2}^{\sigma}Z_{21}^{-1} =
Z_{12}^{-1}L_{2}^{\sigma}Z_{12}L_{1}^{\epsilon}{\cal R}_{21}
\end{equation}
or equivalently
\begin{equation}
{R}_{21}Z_{21}^{-1}L_{1}^{\epsilon}Z_{21}L_{2}^{\sigma}
= Z_{12}^{-1}L_{2}^{\sigma}Z_{12}L_{1}^{\epsilon}{R}_{21}
\end{equation}

\section{Solutions of the Yang--Baxter system}

The problem that we have to  solve  for  determination  of  a
quantum braided group is to find solutions of the
system (\ref{ybez},\ref{yber},\ref{rzz},\ref{zzr}).

 There are several simple
solutions of the system.
One of them is $Z=1, R$ - any solution of the YBE. This gives the
algebras that correspond to the  ordinary  (unbraided)  quantum  groups
 \cite{FRT}. Other solutions are  $Z=R$  or  $Z=PR^{-1}P,\  R$
being any  solution  of  the  YBE. They correspond    to    the
(unquantised)    braided groups.

There are also  trivial  solutions
in the sense that they give no  relations  for  the  algebra
 (\ref{dbgalgebra}).
\\ {\bf Lemma 1}: If  $Z$  is a multiple of the  permutation  matrix P  i.e.
$Z_{ij}^{kl} = z\delta_i^l \delta_j^k$
then     the      only      solution      of      the      system
(\ref{ybez},\ref{yber},\ref{rzz},\ref{zzr})  is   such   that
$R$ is a  multiple   of   the   permutation   matrix   and   the
relation
(\ref{qbgreln})  is  identity. \\
Proof: $P$ solves (\ref{ybez}) and from
(\ref{rzz})  and
(\ref{zzr}) we  get  $R_{ij}  ^{kl}  \delta_n^m  =
R_{jn}^{mk} \delta_i^l$,  solution  of  what  is   a  multiple  of
the permutation matrix $R=rP$.
The relations (\ref{qbgreln}) then read
\[
T_{1}P_{12}T_{2}P_{12}=P_{12}T_{2}P_{12}T_{1}
\]
that is identity.
\\  {\bf  Lemma  2}:  If  $R=rP$  then   an   arbitrary   YBE
solution  $ Z$
solves the system (\ref{ybez},\ref{yber},\ref{rzz},\ref{zzr}) and there are no
 relations in the algebra (\ref{dbgalgebra}).
\\
Proof: The matrix $rP$ solves (\ref{yber}) and the equations (\ref{zzr}) and
(\ref{rzz}) for $R=  P$ are identities
so that any solution of (\ref{ybez}) solves the system. The
relations (\ref{qbgreln}) read
\begin{equation}
P_{12}Z^{-1}_{12}
T_{1}Z_{12}T_{2} =Z^{-1}_{21}T_2Z_{21}T_{1}P_{12} \label{id2}
\end{equation}
that is also identity.

Question is whether there are  solutions   of   the   system
(\ref{ybez},\ref{yber},\ref{rzz},\ref{zzr})  that
provide  nontrivial  examples  of  quantum  braided groups.  The  answer  is
positive even though it seems that they are relatively rare.  The
point is that we know a lot of solutions of the YBE  at  present  but
only a few of their pairs satisfy (\ref{rzz}), (\ref{zzr}).

To present some nontrivial examples
we     are      going      to      solve      the      system
(\ref{ybez},\ref{yber},\ref{rzz},\ref{zzr}) for $n = 2$  i.e.
for  matrices
$R$ and $Z$ of the dimension $4\times 4$. In  this  dimension  we
have  at our disposal  the  complete    list    of    the    YBE
solutions
 \cite{Hietarinta} so that in principle it is easy  to
check  whether  pairs
of the solutions satisfy (\ref{rzz}) and (\ref{zzr}).

There are  two
obstacles  when we  want  to   solve   the system  completely
at least in this dimension.
First, the solutions of the YBE are rather too many. Even  if  we
restrict ourselves to  the  invertible  ones  that  form  eleven   classes
\cite{Hlav87,Hlavybe92} they give 121 pairs  and it takes a lot of
time to check them. Second and more important, even if
we  do  that  we  can  anyway  miss   some   solutions.
 The reason is that
the solutions of the YBE are given up to    symmetries
of the YBE but the cartesian product  of  the  group  of
symmetries
is   not   the   group   of      symmetry   of   the   system
(\ref{ybez},\ref{yber},\ref{rzz},\ref{zzr}).  Indeed,
if      $R$      and      $Z$      solve      the      system
(\ref{ybez},\ref{yber},\ref{rzz},\ref{zzr}) matrices
\begin{equation}
R' = (V\otimes V) R (V\otimes V)^{-1}, \ \ \ V\in  SL(2,C),
\label{rtr} \end{equation}
\begin{equation}
S' = (W\otimes W) S (W\otimes W)^{-1}, \ \ \ W\in SL(2,C),
\label{ztr}
\end{equation}
solve  the  YBE  (\ref{ybez})  and   (\ref{yber})   but   not
necessarily  the   equations
(\ref{rzz})
and (\ref{zzr}).

Therefore, to solve the system completely one can  take  just  one
 solution from the symmetry class (\ref{rtr}) but all solutions
from  the  class
 (\ref{ztr}) (or vice versa). This makes  the  inspection  of
solutions  of
the YBE  very  complicated.  Nevertheless  checking  many
solutions of the YBE  we  have  been  able  to  find  several
solutions  of
the system (\ref{ybez},\ref{yber},\ref{rzz},\ref{zzr}) that give
examples of the quantum braided groups in two dimensions.

A part of the inspection can be made analytically:
\\
{\bf Lemma 3}: If $Z$ is a diagonal solution of the YBE, $Z =
diag(x,u,v,y),\ xuvy  \neq  0  $,
then    there    are    three    types    of     solutions     of
the system (\ref{ybez},\ref{yber},\ref{rzz},\ref{zzr}):
\\
1) If $x=u =v=y$ then R is an arbitrary YBE solution.
\\
2) If $x^{2}= u^{2}= v^{2}= y^{2}$ then R is an arbitrary  YBE
 solution of the eight--or--less--vertex form
\begin{equation}
R=
\left( \begin{array}{cccc}
q&0&0&a\\0&r&b&0\\0&c&s&0\\d&0&0&t       \end{array}
\right)
\label{8vf} \end{equation}
\\
3)R is an arbitrary
YBE solution  of  the  six--or--less--vertex  form  i.e.  (\ref{8vf})  where
$a=d=0$.
\\
Proof: Let $Z$  is  diagonal  i.e.  $Z_{ij}  ^{kl}  =  z_{ij}d_{i}
^{k}d_{j} ^{l}$. Then we get from (\ref{rzz})
\begin{equation}
R_{ij} ^{kl}z_{km}z_{lm}=z_{jm}z_{im}R_{ij} ^{kl}
\  \   {\rm no\
summation.  }
\end{equation}
 Similarly from (\ref{zzr}) we get
\begin{equation}
z_{mi}z_{mj}R_{ij}  ^{kl}= R_{ij}  ^{kl}z_{mk}z_{ml}\  \  {\rm no
\ summation.}
\label {cnd1}
\end{equation}
Therefore, if there are  $i,j,k,l,m$  such that
\begin{equation}
z_{km}z_{lm}\neq   z_{jm}z_{im}\ {\rm  or}   \    z_{mi}z_{mj}\neq
z_{mk}z_{ml} \label{zz}
\label{cnd2}
\end{equation}
then $R_{ij} ^{kl}=0$ .
Substituting
$z_{11}=x,
\ z_{12}=u,\ z_{21}=v,\ z_{22}=y$.  into the   condition
(\ref{cnd1},\ref{cnd2})
gives the cases 1) -- 3).

The case 1) of the lemma 3 gives ordinary quantum groups.
The  case   2)   implies    that   only  eight-or-less-vertex
 matrices $R$  can  be  used  for
quantization of supergroups (cf. \cite{Hlavijmp}). The case  3)  will
be discussed in the next Section.

\section{Examples}

The first type of nontrivial examples provides  us  Lemma  3,
case 3). From the list of solutions of  the  YBE  on
can see that
there  are  just  four  six-or-less-vertex invertible
 solutions   of   the   YBE.
 The first one is $R=rP$ that due to the Lemma 2 gives
trivial "no relation" quantum braided group. The other three are
\begin{equation}
R_{5}=
\left( \begin{array}{cccc}
q&0&0&0\\0&1&0&0\\0&q-t&qt&0\\0&0&0&q       \end{array}
\right)                              ,\ \
R_{6}=
\left( \begin{array}{cccc}
q&0&0&0\\0&1&0&0\\0&q-t&qt&0\\0&0&0&-t      \end{array}
\right)                              ,\ \
\end{equation}
\begin{equation}
R_{8}=
\left( \begin{array}{cccc}
q&0&0&0\\0&r&0&0\\0&0&s&0\\0&0&0&t       \end{array}
\right)
\end{equation}
(The  numbering  corresponds  to  that in classifications   given   in
\cite{Hlav87,Hlavybe92}.)

The defining relations
of the quantum braided group  given  by  $R=R_{5},\  Z=diag(x,u,v,y)$
 are
\begin{eqnarray}
   tAB &= & \sigma BA, \nonumber
\\ qCA &= & \sigma AC, \nonumber
\\ qDB &= & \tau BD, \label{qbg58}
\\ tCD &= & \tau DC, \nonumber
\\ \sigma \tau BC &= & qtCB, \nonumber
\\ \sigma (AD-DA) &= & (q-t)CB, \nonumber
\end{eqnarray}
where $\sigma=v/x,\ \tau=y/u$ and $A,B,C,D$ are generators of
the algebra.
\begin{equation}
T=\left( \begin{array}{cc}
A&B\\C&D \end{array}
\right)     .
\end{equation}
This quantum braided group have  a
structure similar to   the    well    known    quantum  group
$GL_{q,s}(2)$
\cite{BHGLqt} which  is
obtained when $y=uv/x$.

On the other hand, the quantum braided group  given  by  $R=R_{6},\
Z=diag(x,u,v,y)$
where $q \neq -t$ reminds the quantum supergroup
$GL_{q,t}(1|1)$ \cite{dab}. The
defining relations are
\begin{eqnarray}
B^2 & = & 0 \ =\ C^2 \nonumber
\\ tAB &= & \sigma BA, \nonumber
\\ qCA &= & \sigma AC, \nonumber
\\ tDB &= & -\tau BD, \label{qbg68}
\\ qCD &= & -\tau DC, \nonumber
\\ \sigma \tau BC &= & qtCB, \nonumber
\\ \sigma (AD-DA) &= & (q-t)CB, \nonumber
\end{eqnarray}

There are several cases of quantum braided groups given by
$R=R_{8}=diag(q,r,s,t)$,
$Z=R_{8}'=diag(x,u,v,y)$ (cf. \cite{hlavqs}).

If $q=t,\  q^{2}=rs$  we
get  six  relations algebra
\begin{eqnarray}
    AB &= & \kappa BA, \nonumber
\\  CA &= & \kappa AC, \nonumber
\\  DB &= & \rho BD, \label{qbg88}
\\  CD &= & \rho  DC, \nonumber
\\  CB &= & \omega BC, \nonumber
\\ AD &= & DA, \nonumber
\end{eqnarray}
where $\kappa=vq/(xs),\ \rho=ry/(ut),\ \omega=rvy/(sxu)$.
When one  of  the  above  relations  between  the  parameters
$q,r,s,t$  does
not hold we get additional relations. Namely if $q\neq t$, then
\begin{equation}
B^{2}=0=C^{2}.
\end{equation}
If $q^{2}\neq rs$ then
\begin{equation}
AB=BA=AC=CA=0.
\end{equation}
If $t^{2}\neq rs$ then
\begin{equation}
BD=DC=BD=DB=0.
\end{equation}
In the generic case when $q \neq t,\ q^{2} \neq  rs  \neq  t^{2}$,
the quantum braided group is defined by
\[ AB=BA=AC=CA=BD=DC=BD=DB=B^{2}=C^{2}=0, \]
\begin{equation}
 AD=DA,\ CB = \omega BC.
\end{equation}

The braiding  relations  for  all  the  above  given  quantum
braided groups are  given  by
matrix $Z=R'_8=diag(x,u,v,y)$ and read
\begin{eqnarray}
\psi (A \otimes X) = X \otimes A,\ \ \psi (X \otimes A) &=& A
\otimes X,\  \
X\in \{ A,B,C,D \}, \nonumber
\\   \psi (B \otimes B) & = & \xi B \otimes B,\nonumber
\\ \psi (B \otimes C) & = & \xi^{-1} C \otimes B,
\\ \psi (C \otimes B) & = & \xi^{-1} B \otimes C,\nonumber
\\ \psi (C \otimes C) & = & \xi C \otimes C,\nonumber
\\ \psi (D \otimes X) = X \otimes D,\ \ \psi (X \otimes D) & =&  D
\otimes X,\  \
X\in \{ A,B,C,D \} \nonumber
\end{eqnarray}
where $\xi  = xy/(uv)$. Note that $A,\ D $ are always bosonic.
Only $B$ and $C$ can have anomalous statistics
for this $Z$.

Other   examples   provide   the   solutions   of   the    system
(\ref{ybez},\ref{yber},\ref{rzz},\ref{zzr}) where
\begin{equation}
Z=R'_{10} =
\left( \begin{array}{cccc}
1&0&0&0\\x&1&0&0\\y&0&1&0\\z&y&x&1       \end{array}
\right)
 \end{equation}
The braiding $\psi$ given by this matrix is
\begin{eqnarray}
\psi (C \otimes X)& =& X \otimes C,\ \ \psi (X \otimes C) = C \otimes X,\
\ X \in \{A,B,C,D\}, \nonumber
\\ \psi (A \otimes A) &=&  A  \otimes  A  +  \tau  C\otimes  C,
\nonumber
\\ \psi (A \otimes B) &=& B \otimes A - \tau(A-D)\otimes  C
,\nonumber
\\ \psi (A \otimes D) &=& D \otimes  A  -  \tau  C\otimes  C,
\nonumber
\\ \psi (B \otimes A) &=& A \otimes B - \tau  C\otimes (A-D),
\\ \psi (B \otimes B) &=& B \otimes  B  +  \tau  (A-D)\otimes
(A-D)+2\tau^{2} C\otimes C, \nonumber
\\ \psi (B \otimes D) &=& D \otimes B + \tau  C\otimes (A-D), \nonumber
\\ \psi (D \otimes A)&=&  A  \otimes  D  -  \tau  C\otimes  C
,\nonumber
\\ \psi (D \otimes B)&=& B  \otimes  D  +  \tau (A-D)\otimes  C
, \nonumber
\\ \psi (D \otimes D)&=&  D  \otimes  D  +  \tau  C\otimes  C
,\nonumber
\end{eqnarray}

where $\tau = z-xy$. Note that there are  again  two  bosonic
elements, namely $C$ and $A+D$. For  $z=xy$  the  braiding
 is bosonic even though $Z\neq 1$.

There     are     two     solutions     of     the     system
(\ref{ybez},\ref{yber},\ref{rzz},\ref{zzr}) with  $Z=R'_{10}$.  The
 matrix $R$ then is either
\begin{equation}
R=R_{10}=
\left( \begin{array}{cccc}
1&0&0&0\\g&1&0&0\\h&0&1&0\\f&h&g&1       \end{array}
\right)\ \ {\rm or} \ \
R=R_{11} =
\left( \begin{array}{cccc}
1&0&0&0\\-g&1&0&0\\g&0&1&0\\-gh&h&-h&1       \end{array}
\right),
\end{equation}

The quantum braided group given by $Z=R'_{10},\ R=R_{11}$ is
defined by the relations
\begin{eqnarray}
BA &=& AB + \chi B^{2}, \nonumber
\\ DB &=& BD + \gamma B^{2},\nonumber
\\ CB &=& BC + \gamma AB- \chi BD,
\\ CA &=& AC + \tau AB  + \gamma A(A-D)+ \gamma BC +
\tau (\chi - \gamma)B^{2} - \gamma \chi BD, \nonumber
\\ DC &=& CD - \chi (A-D)D + \chi BC - \tau (\chi -\gamma )
B^2 + (\tau-\chi^2)BD, \nonumber
\\ DA &=& AD + \gamma AB + \chi BD + (
\tau + \gamma  \chi)B^{2} ,\nonumber
\end{eqnarray}
where $\tau =z-xy,\  \chi =y-h,\ \gamma =y-g$. Note that for
$\tau =0$ we get the
nonstandard unbraided deformation          of          $GL(2)$
\cite{Zakrzewski,EOW} even if $Z \neq 1$.

The last example of the quantum braided group in this paper is
given  by  $R=R_{10},\
Z=R_{10}'$. We can assume that the parameters of $R_{10}$ satisfy $g+h
\neq  0$  or  $
f  \neq  gh$ because otherwise we get a special case of the previous example.
Under this assumption the defining relations read
\[ AB=BA=DB=BD=B^{2}=0, \]
\[ AD =DA,\ BC=CB,\ A^{2}=D^{2}, \]
\begin{equation} CA = AC + (y-h)A(A-D) + (y+g)BC
\end{equation}
\[ CD = DC + (y-h)D(A-D) - (y+g)BC \]
If $g+h \neq 0$ then moreover
\begin{equation}
BC=A(A-D),\ \ \ \ AC=CD.
\end{equation}

\section{Conclusions}

We have written down the defining relations of a new type of algebras
that generalize both the quantum groups and braided groups as
well as the quantum supergroups. The
relations of the algebras are determined by a pair of
matrices $(R,Z)$  that solve a
 system of Yang--Baxter--type equations.
The algebras can be extended to bialgebras with matrix coproduct and
counit, however, the multiplication in the tensor product of the
algebras is defined by virtue of the braiding map given by the
 matrix $Z$.

Besides   simple   solutions   of   the    system   of    the
Yang--Baxter--type equations
that generate either quantum groups or braided groups, we have found
several solutions
 that generate genuine quantum braided groups in the sense that by a
choice of parameters we can get quantum groups or braided groups
as a special cases.

This work was supported in part by the grant CSAV No. 11 086.

\end{document}